\documentclass[prl,twocolumn,superscriptaddress,noshowpacs]{revtex4}
\usepackage{amsmath}
\usepackage{graphicx}
\usepackage{bm}
\usepackage[T1]{fontenc}
\usepackage[latin1]{inputenc}
\usepackage{times}
\usepackage{amssymb}
\usepackage{color}
\usepackage{units}
\usepackage{ulem}

\newcommand{\sqbr}[1]{\left[ #1 \right]}
\newcommand{\modbr}[1]{\left| #1 \right|}
\newcommand{\etal}{\textit{et al.~}}

\begin{document}
\title{Spatio-temporal focussing of an ultrafast pulse through a multiply scattering medium}

\author{David J.\ McCabe}
\author{Ayhan Tajalli}
\affiliation{ CNRS-Universit\'e de Toulouse, UPS, Laboratoire Collisions, Agr\'egats R\'eactivit\'e,
IRSAMC, F-31062 Toulouse, France}
\author{Dane R.\ Austin}
\affiliation{Clarendon Laboratory, Department of Physics, University of Oxford, Oxford, OX1 3PU, United Kingdom}
\affiliation{ICFO-Institut de Ciencies Fotoniques, Mediterranean Technology Park, 08860 Castelldefels (Barcelona), Spain}
\author{Pierre Bondareff}
\affiliation{Institut Langevin, ESPCI ParisTech, CNRS UMR 7587, Universit\'es Paris 6 \& 7, INSERM, ESPCI, 10 rue Vauquelin, Paris, 75005, France}
\author{Ian A.\ Walmsley}
\affiliation{Clarendon Laboratory, Department of Physics, University of Oxford, Oxford, OX1 3PU, United Kingdom}
\author{Sylvain Gigan}
\affiliation{Institut Langevin, ESPCI ParisTech, CNRS UMR 7587, Universit\'es Paris 6 \& 7, INSERM, ESPCI, 10 rue Vauquelin, Paris, 75005, France}
\author{B\'{e}atrice Chatel}
\email{beatrice.chatel@irsamc.ups-tlse.fr}
\affiliation{ CNRS-Universit\'e de Toulouse, UPS, Laboratoire Collisions, Agr\'egats R\'eactivit\'e,
IRSAMC, F-31062 Toulouse, France}

\begin{abstract}
Pulses of light propagating through multiply scattering media undergo complex spatial and temporal distortions to form the familiar speckle pattern. There is much current interest in both the fundamental properties of speckles and the challenge of spatially and temporally refocussing behind scattering media. Here we report on the spatially and temporally resolved measurement of a speckle field produced by the propagation of an ultrafast optical pulse through a thick strongly scattering medium. By shaping the temporal profile of the pulse using a spectral phase filter, we demonstrate the spatially localized temporal recompression of the output speckle to the Fourier-limit duration, offering an optical analogue to time-reversal experiments in the acoustic regime. This approach shows that a multiply scattering medium can be put to profit for light manipulation at the femtosecond scale, and has a diverse range of potential applications that includes quantum control, biological imaging and photonics.
\end{abstract}

\maketitle

\section{Introduction}

The multiple scattering of coherent light is a problem of both fundamental and applied importance \cite{Sebbah2001}. In optics, phase conjugation allows spatial focussing and imaging through a multiply scattering medium \cite{Vellekoop2007, Popoff2010, Yaqoob2008}; however, temporal control is nonetheless elusive \cite{DelaCruz2004}, and multiple scattering remains a challenge for femtosecond science. When light propagates through a thick multiply scattering medium, the large number of scattering events exponentially deplete the ballistic photons with propagation distance, and the transmitted light is dispersed in a highly complex manner. When illuminated by a single-frequency laser, such a medium is known to give rise to a spatial speckle \cite{Goodman1976}. Furthermore, this speckle is wavelength-dependent:  illuminated by a broadband laser, the medium therefore also produces a spectral speckle whose characteristic size is inversely related to the confinement time in the medium, or Thouless time. It corresponds to a severe temporal broadening of the pulse \cite{Patterson1989,Tomita1995}. Nonetheless, multiple scattering is deterministic and coherence is not destroyed. In the spatial domain, recent experiments have thus demonstrated that wavefront shaping is able to generate an intense focus through an opaque medium, and can be interpreted as phase conjugation \cite{Vellekoop2007, Popoff2010,Yaqoob2008}. For the acoustic and GHz-electromagnetic regimes, time-reversal experiments are the counterpoint to this principle in the time domain \cite{Fink1997, Derode2001, Lerosey2007}: a short pulse incident on a medium generates a long coda, which may be time-reversed and retransmitted through the medium in order to regain the initial short pulse. These experiments have not only demonstrated that multiple scattering can be compensated but --- more importantly --- that this can actually improve addressing, imaging or communication \cite{Derode2003, Vellekoop2010}. Due in particular to the inability to measure electric fields directly in the temporal domain at higher frequencies, an optical domain time-reversal experiment remains elusive; yet the ability to measure and shape femtosecond electric fields in the spectral domain nevertheless offers the opportunity of a route to the same goal.

From a quantum-control perspective, the use of optimally shaped pulses to control light-matter interactions \cite{Rabitz2000} has been a fertile field of research \cite{Assion1998,Silberberg2009,Weiner2000,Monmayrant2010}. The absence of predesigned control mechanisms in complex systems may be obviated by a closed-loop feedback scheme \cite{Warren1993} to find efficient pulse shapes at the expense of insight into the physical mechanism \cite{Assion1998}. By contrast, theoretically tractable model systems permit an open-loop approach where the appropriate control pulse is calculated directly \cite{Dudovich2002,Degert2002}. In parallel, the propagation of ultrafast pulses through strongly dispersive media has been a topic of research for many years. Atmospheric aberrations have been successfully compensated \cite{Ackermann2006} as well as the dispersion due to thick optical media \cite{Delagnes2007}. However, despite some insight into the multiple scattering of ultrafast pulses \cite{Wen-Jun2010,Calba2008,Small2009}, measurement of the scattering medium transfer function at a single speckle grain via scanning the frequency of a c.w.\ laser \cite{Webster2004,Gerke2005}, and some two-photon excitation experiments in scattering samples \cite{DelaCruz2004}, the effects of multiple scattering have ultimately been considered too complicated to be compensated.

In this paper, we initially demonstrate a full spatio-temporal characterization of a femtosecond speckle field using spatially and spectrally resolved Fourier-transform interferometry (SSI).  Due to the linearity of the scattering process, knowledge of the spectral phase facilitates active temporal focussing of the speckle via the open-loop feedback of the measured phase to a spectral pulse shaper placed before the sample. Here we give the first experimental demonstration of this effect. This experiment relies on a spatially resolved phase measurement, since the lack of large-scale spatial homogeneity in the speckle field prevents the spatial integration typical of control experiments. An effective phase compensation is spatially localized according to the correlation length; thus a spatially resolved spectral shaper is not a prerequisite for a degree of spatial control.

\section{Results}

\begin{figure}
\centering
\includegraphics [width = \columnwidth]{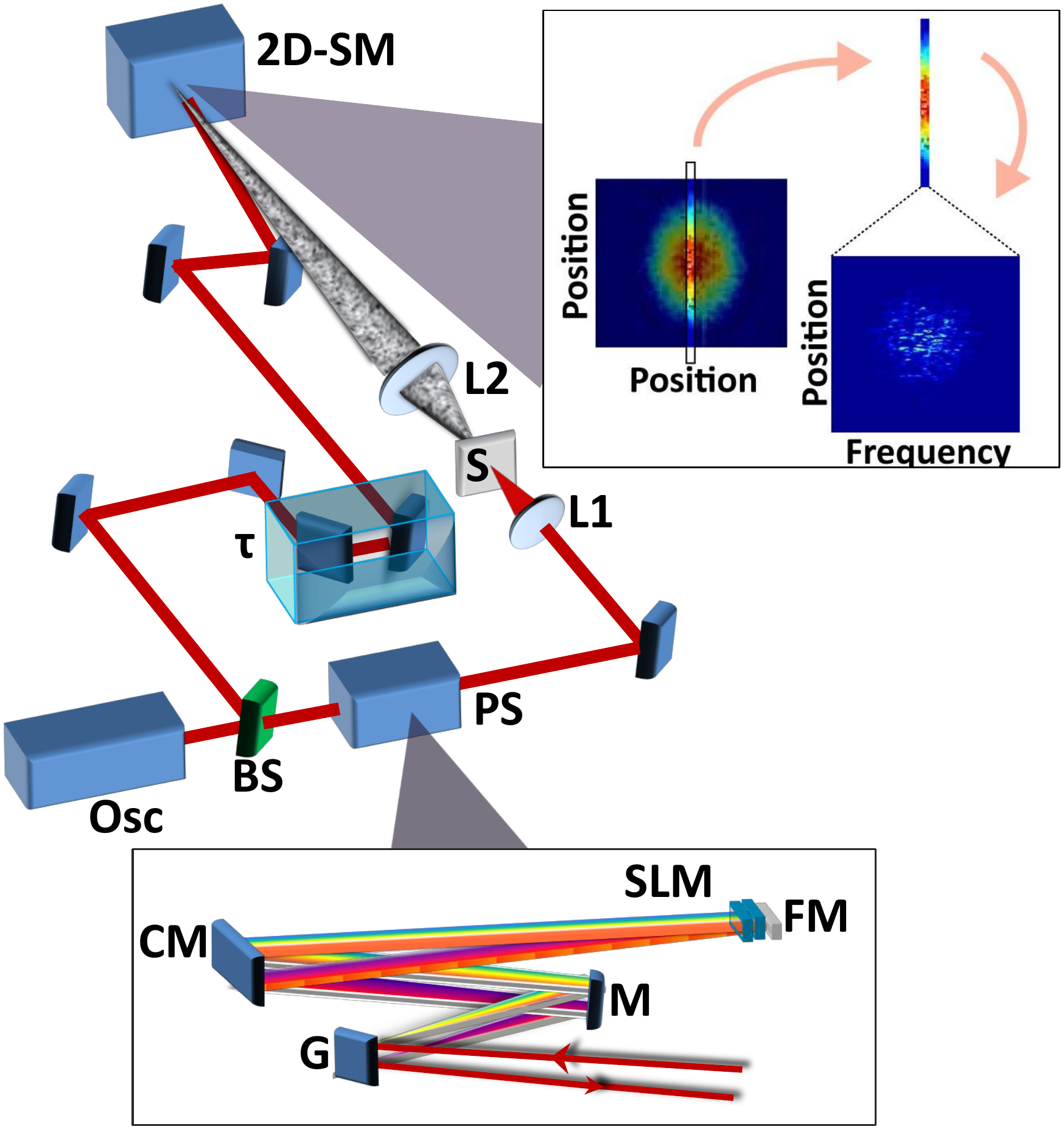}
\caption{The experimental set-up. The output of a laser oscillator (Osc) is divided by a beamsplitter (BS) to form an interferometer. One arm passes through a pulse shaper (PS) and is focussed onto the sample (S) by lens L1. The output speckle field is imaged by lens L2 onto the spectrometer slit (2D-SM), which performs a measurement of the spectral intensity spatially resolved along the slit (upper inset). The reference arm is combined with an adjustable delay $\tau$ and angle. The spectral pulse shaper is a folded $4f$ line, and comprises a grating (G), cylindrical mirror (CM), plane mirror (M), folding mirror (FM) and spatial light modulator (SLM).}
\label{fig:setup}
\end{figure}

\subsection{Spatiotemporal characterization of the speckle field}

\begin{figure*}
\centering
\includegraphics [width = \textwidth]{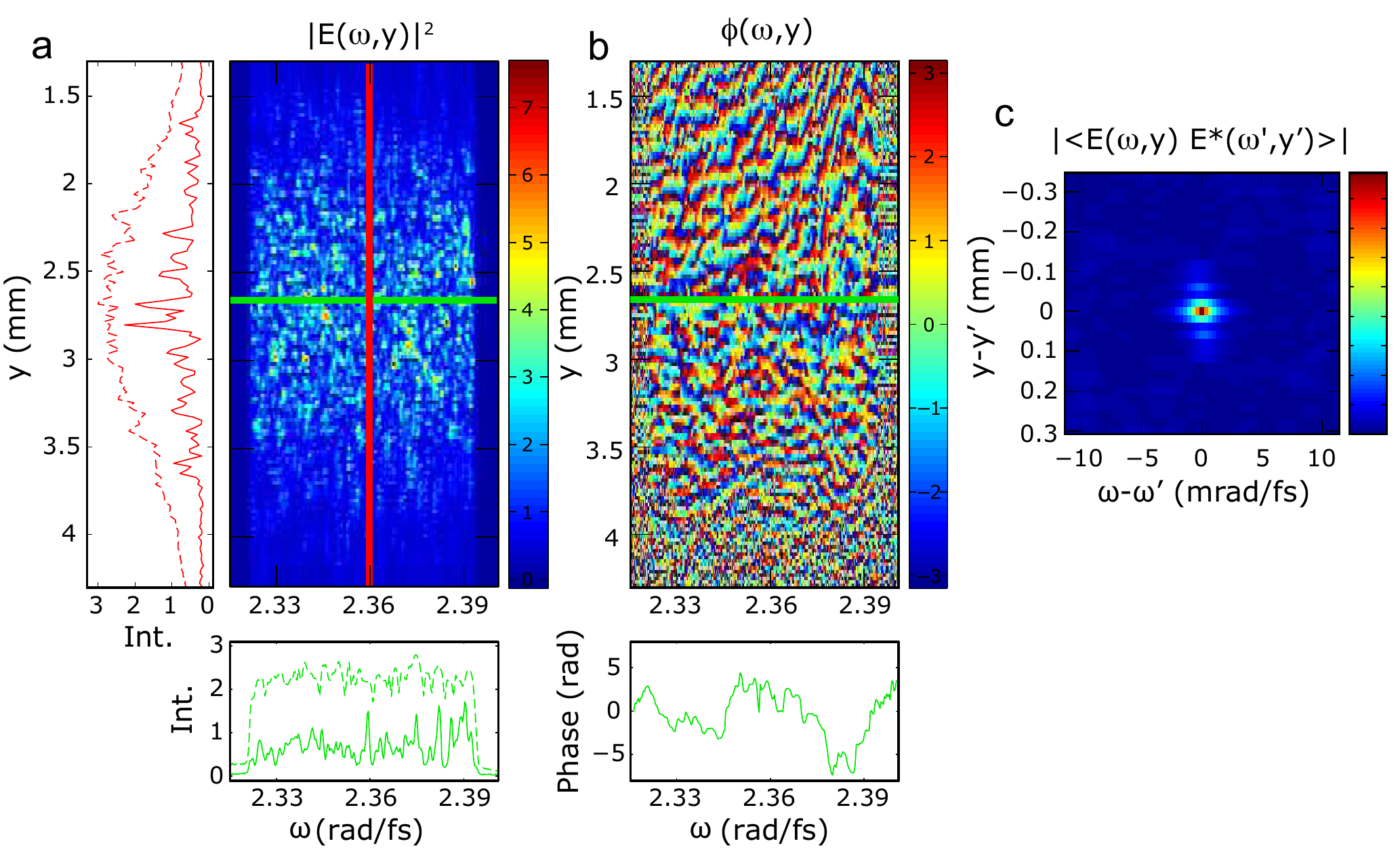}
\caption{Experimental reconstruction of the spatio-spectral speckle electric field $E_\textrm{out}(\omega, y)$ after propagation through a multiply scattering medium. (a) Speckle intensity $\modbr{E_\textrm{out}(\omega, y)}^2$, with respective projections onto the spatial (red) and spectral (green) axes --- individual slices (solid lines) and integrated signals (dashed lines) are considered. (b) Speckle phase $\phi_\textrm{out}(\omega, y)$, along with a spectral phase measurement $\phi_\textrm{out}(\omega, y_0)$ localized along a single spatial slice. (c) Autocorrelation function $\left| \left< E_\textrm{out}(\omega, y) E_\textrm{out}^{*}(\omega', y')\right> \right|$, demonstrating that the speckle is well resolved both spatially and spectrally. The spatial correlation length is related to the speckle grain size, while the spectral correlation length is the bandwidth of the medium, inversely proportional to the Thouless time.}
\label{fig:measurements}
\end{figure*}

The schematic experimental layout is shown in Fig.\ 1 
(see Methods for details). In essence, an ultrashort pulse passes through a pulse shaper and is focussed on an opaque sample.  The sample is a scattering ZnO layer that transforms the ultrashort pulse into a complex spatio-temporal speckle.  This speckle is stationary on a timescale much longer than the experiment duration. The spatio-temporal speckle is reimaged and spatio-spectrally characterized in phase and amplitude via SSI (see Methods).

Figure 2 
 shows the (a) intensity and (b) phase of a typical spatio-spectral speckle field $E_\textrm{out}(\omega, y)$ as measured by SSI. The spatio-spectral speckle is clearly demonstrated and the complex structure of $E_\textrm{out}(\omega, y)$ is fully resolved in both phase and intensity --- a prerequisite for quantum control. This structure is also visible in the one-dimensional spatial [(a), red, solid] and spectral [(a), green, solid] `lineout' slices indicated. The integrated projections of the speckle intensity [(a), dotted lines], however, show that a spectrally unresolved speckle image, as measured on a camera, would yield a strongly reduced contrast, while a non-imaging spectral measurement would only yield the initial source spectrum; this further motivates the necessity of a spatially resolved phase measurement for the temporal focussing experiment described below. Meanwhile the spectral phase reveals a similarly complex structure (Fig.\ 2b) 
; as a consequence it is clear that a spatially averaged phase measurement offers no utility whatsoever for pulse recompression.

A Fourier transform of the complex field along the spectral axis gives the spatially resolved temporal behaviour of the speckle $E_\textrm{out}(t,y)$ (see Fig.\ 3a). 
 This spatio-temporal field exhibits the same complex speckle structure as before, as evinced by both the 3D plot and the spatial (red) and temporal (black) lineouts projected onto the walls. The spatially (black) and temporally (red) integrated fields are plotted above; the former reveals the confinement time which is fitted as approximately \unit[2.5]{ps}, in good agreement with the spectral bandwidth measured from the autocorrelation function (Fig.\ 2c). 
  For this function, the spatial and spectral correlation distances at the spectrometer were \unit[50]{$\mu$m} (corresponding to \unit[3.6]{$\mu$m} in the object plane) and \unit[2.55]{mrad/fs} respectively. As a further remark, with an additional lateral scan, a full tridimensional speckle measurement could be envisioned. This would be interesting for fundamental studies of speckle properties such as vortices and singularities \cite{OHolleran2009}.

\subsection{Focussing of the speckle field via spectral pulse shaping}

Multiple scattering is a complex yet linear process, therefore the evolution of the pulse electric field through the medium may be elucidated by a Green function formalism. For a given spectral component, and for a distribution of sources  $E_S(r',\omega)$, the resulting field  reads $E(\omega,r)=\iiint G_\textrm{scatt}(\omega, r,r') E_S(\omega, r') d^3r'$, where $G_\textrm{scatt}(\omega,r,r')$ is the Fourier component of the Green function between point $r$ and $r'$, in the presence of scattering. In our simple case where we enter with a focussed beam commensurate with the correlation distance of the scatterers $E_\textrm{in}(\omega)$, the Green function reduces to $E_\textrm{out}(\omega, y)=H_\textrm{scatt}(\omega,y)H_\textrm{shaper}(\omega) E_\textrm{in}(\omega)$, where $H_\textrm{scatt}(\omega, y) = A_\textrm{scatt}(\omega, y) \exp [i \phi_\textrm{scatt}(\omega, y)]$ and $H_\textrm{shaper}(\omega) = A_\textrm{shaper}(\omega) \exp [i \phi_\textrm{shaper}(\omega)]$ are the medium and shaper transfer functions respectively, with both consisting of phase and amplitude contributions. For the measurements presented above, the SSI measurement reveals the relative phase $\phi_\textrm{scatt}(\omega, y) - \phi_\textrm{in}(\omega)$ with the pulse shaper set to $H_\textrm{shaper}(\omega)=1$. The coupling between spatial and spectral modes in the scatterer, and the concomitant inseparability of $H_\textrm{scatt}(\omega, y)$ into a product of spatial and spectral components, is the root of the exigence of a spatially resolved measurement for a spectral phase correction. This coupling has very recently been exploited to exert a degree of temporal control over isolated speckle grains via shaping of the spatial mode \cite{Aulbach2010,Katz2010}. In general, a complete reversal of the scattering process would therefore require a spatio-spectrally resolved shaping element covering all the contributing modes, rather than successive shapers controlling the spatial and spectral degrees of freedom independently.

\begin{figure*}
\centering
\includegraphics [width = \textwidth]{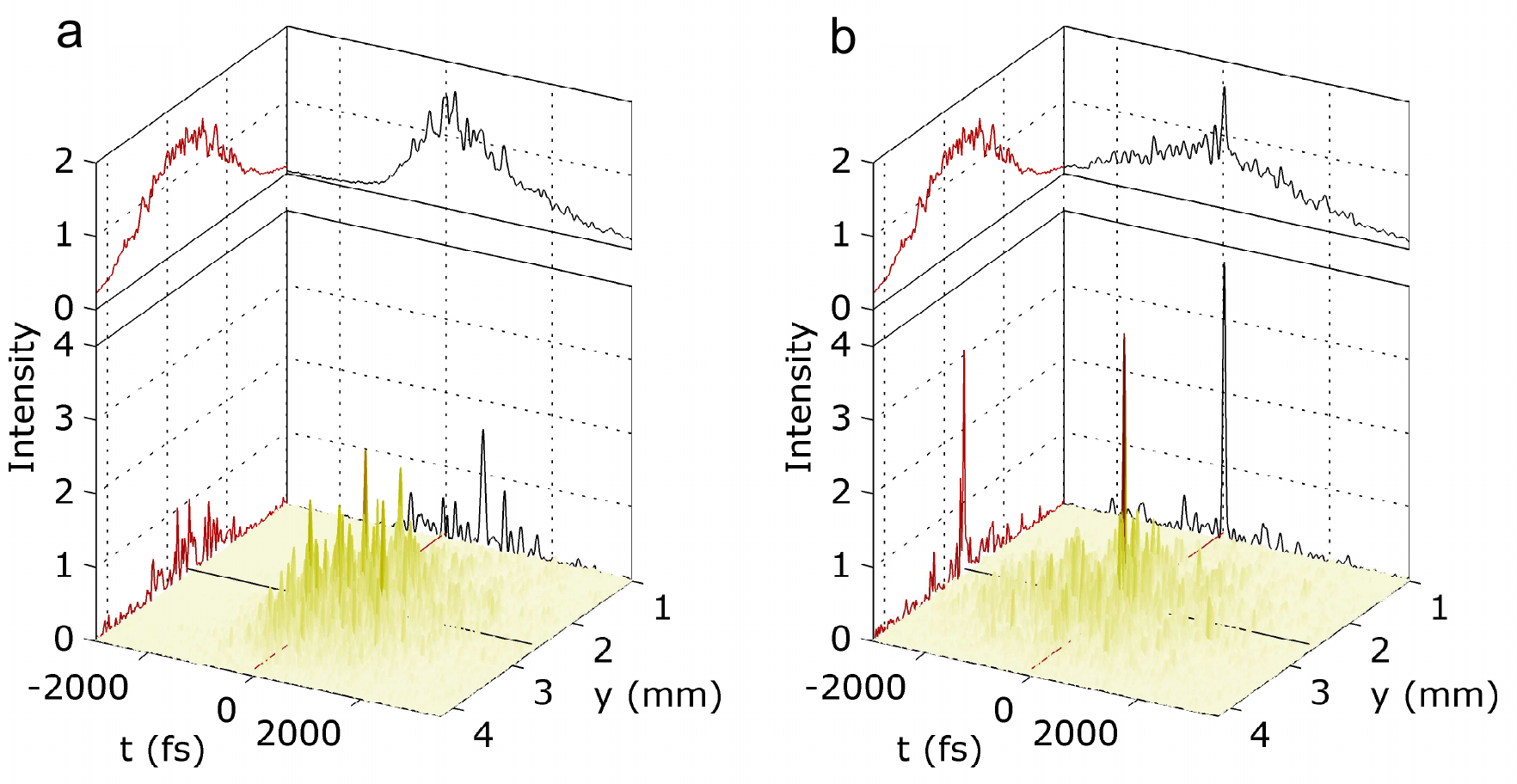}
\caption{Reconstructed spatio-temporal intensities $\modbr{E_\textrm{out}(t, y)}^2$ (a) before and (b) after compensation of the phase at \unit[$y_0=2.66$]{mm}. An intense peak emerges from the background with a contrast ratio of 15. One-dimensional `lineouts' at the location of this peak (projections onto walls) and integrated signal (top) show that the peak is focussed in time (black) and localized in space (red). The temporal and spatial widths of the peak are the Fourier-limit pulse duration and the spatial phase correlation distance respectively.}
\label{fig:temporal-focus}
\end{figure*}

In a subsequent experiment, such an open-loop compensation of the spectral phase is performed. A position $y_0$ is selected and the transfer function $\phi_\textrm{shaper}(\omega) = - \phi_\textrm{scatt}(\omega, y_0)$ is applied to the pulse shaper. A successful phase compensation at $y_0$ is demonstrated by a further SSI measurement of \mbox{$\phi_\textrm{out}(\omega, y_0) - \phi_\textrm{in}(\omega, y_0) = 0$}; furthermore, this flat output phase (and consequent temporal focus) is spatially localized as determined by the spatial phase correlation length. (It should be noted that phase-only shaping is sufficient to obtain a near--transform-limited pulse duration; amplitude shaping would serve predominantly to reduce the background in the temporal domain \cite{Derode1999}).

\section{Discussion}
Experimentally, the shaped pulse is found to have a full width at half-maximum duration of \unit[59]{fs} (close to the transform-limited duration of \unit[54]{fs}); the optimal phase is reached after two iterations of the feedback procedure. In the temporal domain, this corresponds to an intense temporally focussed and spatially localized pulse that emerges from the background spatio-temporal speckle (Fig.\ 3b). 
 The peak temporal intensity has a contrast ratio of 15 relative to the average background before compensation, and the average background along $y_0$ is reduced by a factor of 2. As shown by the projected `lineout' intensities, the temporal focus is spatially localized; the localization distance is \unit[30]{$\mu$m} which is commensurate with the correlation distance of Fig.\ 2c. 
  The spatially integrated temporal field (Fig.\ 3
, top, black), exhibits the redistribution of the temporal intensity after shaping that is the signature of a temporal focus. The temporally integrated field, however, is not altered (Fig.\ 3 
, top, red): the peak can be said to be spatially localized rather than focussed. Due to the spatially localized nature of the phase compensation, the peak integrated temporal intensity is not significantly altered. This ability to temporally recompress the pulse without spatial redistribution of energy may have important applications for the nonlinear imaging of biological samples, where a rise in fluence may result in the onset of thermally induced damage. As shown in time reversal and wavefront shaping experiments, another striking feature is that the more scattering the medium is, the more efficient the focussing will be. Indeed, the signal-to-noise ratio is governed by the ratio of the Thouless time of the medium over the initial duration of the pulse, i.e.\ the number of independent spectral degrees of freedom.  Furthermore, in contrast with conventional phase compensation techniques, here all shaping imperfections affect the signal-to-noise ratio; the temporal duration and spatial localization are limited by the Fourier-limit duration and autocorrelation size of the speckle respectively \cite{Vellekoop2010}.

In conclusion, we have characterized the speckle spatio-temporal electric field of a multiply scattered ultrafast pulse. Furthermore, we have implemented an open-loop correction of the spectral phase in order to produce a naturally Fourier- and diffraction-limited pulse after the medium at a chosen position. The results demonstrate the benefits of a spatially resolved measurement: typical control experiments consider a spatial average, but for such systems that lack large-scale correlations such averaged measurements are valueless. Moreover, we have demonstrated that these correlations permit spatial control without spatial resolution in the spectral pulse-shaper. Our study branches out previous spatial-speckle shaping techniques to the temporal domain and bridges the gap with time-reversal experiments in acoustics and electromagnetism. This capacity to recover a short pulse on a selected spatial speckle point has important potential applications for quantum control and photonics, as well as for the fundamental studies of complex media. It suggests that ultrafast diagnostic techniques including nonlinear microscopy and time-resolved spectroscopy may be performed deep within or beyond biological tissue, beyond the range of ballistic photons.

\section{Methods}

\subsection*{Experimental system}
The ultrafast source used for this experiment is an \unit[80]{MHz} oscillator that delivers \unit[4]{nJ} pulses at \unit[800]{nm} with a spectral bandwidth of \unit[80]{nm}. The oscillator pulse train is divided to form the two arms of an interferometer. In one arm is placed both a spectral pulse shaper and the scattering sample; the other arm acts as a reference and is recombined with an adjustable relative angle and delay at the entrance slit of a home-built imaging spectrometer \cite{Austin2009}. The spectral calibration was performed using a calibration lamp and verified against a commercial spectrometer; to ensure consistency the same calibration was used for the pulse shaper. The spectrometer response --- calibrated against a white-light source --- varied by less than 10\% over the bandwidth of the pulse. The spectrometer has spatial and spectral optical resolutions of \unit[20]{$\mu$m} and \unit[0.4]{nm} respectively; the two-dimensional detector permits the speckle spectrum to be spatially resolved along the slice that falls on the entrance slit (see inset of Fig.\ 1). The spatio-spectral intensity adhered to the well-known exponential decay probability distribution \cite{Goodman1976}, confirming that the speckle pattern was fully resolved. For a typical measurement $8 \times 10^5$ oscillator pulses are integrated.

The phase and amplitude pulse-shaper comprises a pair of liquid-crystal spatial light modulators (SLMs) in the Fourier plane of a folded double-pass $4f$ zero-dispersion line \cite{Monmayrant2004}. In order to optimize the trade-off between bandwidth and shaping resolution, the $4f$-line grating is chosen so as to overfill the SLMs; thus the output bears a clipped \unit[30]{nm} hyper-gaussian spectrum and may be shaped arbitrarily within a \unit[23]{ps} time window. This output is then focussed to a waist of approximately \unit[15]{$\mu$m} onto the surface of the sample. The scattered, transmitted light is collected by a lens with a numerical aperture of 0.25 and imaged onto the spectrometer slit with a magnification factor of 14.

\subsection*{SSI Technique} The SSI technique \cite{Tanabe2002,Monmayrant2010} performs a relative measurement of the spectral phase between a reference and unknown pulse. The interference between the sample image and reference beams causes interference fringes on the spectrometer whose spectral and spatial periods are determined by the relative delay and angle of the two beams respectively. These fringes are additionally modulated by the relative phase between the two beams.

\begin{align}
\label{eq:SI}
S(x,\omega) & = \modbr{A_\textrm{s}(\omega,y) e^{i\phi_\textrm{s}(\omega,y)} + A_\textrm{r}(\omega,y) e^{i\sqbr{\phi_\textrm{r}(\omega,y) + \omega \tau + k_y y}}}^2 \notag \\
& = \modbr{A_\textrm{s}(\omega,y)}^2 + \modbr{A_\textrm{r}(\omega,y)}^2 \notag \\
& + \modbr{A_\textrm{s}(\omega,y)} \modbr{A_\textrm{r}(\omega,y)} \cos \sqbr{\phi_\textrm{s}(\omega,y) - \phi_\textrm{r}(\omega,y) - \omega \tau - k_y y}.
\end{align}

Here $\tau$ is the time delay between the two pulses and $k_y$ is the difference between the transverse components of the propagation vectors (such that their subtended angle is $\theta = k_y/\modbr{\mathbf{k}}$). $A_\textrm{s}$, $A_\textrm{r}$, $\phi_\textrm{s}$ and $\phi_\textrm{r}$ denote the spatio-spectral amplitude and phase of the scattered (s) and reference (r) pulse respectively.

In order to recover the amplitude and phase of the unknown pulse, the raw interferogram is thus Fourier transformed along both the spatial and spectral axes, and the relative spatio-spectral phase may be extracted through the filtering of an a.c.\ term followed by the inverse Fourier transform. This isolates one of the complex exponential terms that correspond to the cosine of the final summand of Eq.\ \ref{eq:SI}; the argument of this exponential reveals the relative spectral phase modulo $\pm \pi$.

\subsection*{Sample preparation}
The scattering medium is thick layer of ZnO powder (a widely used white paint component) deposited homogeneously on a microscope slide by sedimentation. The thickness has been measured to be \unit[$L=35 \pm 5$] {$\mu$m}, the transport mean free path has been measured as \unit[$l^*=2.1 \pm 0.2$]{$\mu$m}, and absorption is known to be negligible. Since $L > 10 l^*$, the multiple scattering regime applies and virtually no ballistic light traverses the medium.

\clearpage

\bibliographystyle{DJMbibstyle}
\bibliographystyle{osajnl}


\vspace{0.8cm}
\textbf{Acknowledgements}
\vspace{0.5cm}	

We acknowledge N. Curry and S. Gr\'{e}sillon for sample preparation and characterization as well as useful discussions, A.Wyatt for fruitful discussions on  the characterization and E. Baynard and S. Faure for technical assistance. This work was supported by Agence Nationale de la Recherche via ANR COCOMOUV, Marie Curie Initial Training Network (grant CA-ITN-214962-FASTQUAST) and Alliance (Partenariats Hubert Curien/British Council). P.B.\ is funded by ANR ROCOCO. I.A.W.\ acknowledges support by the Royal Society.

\vspace{0.8cm}
\textbf{Author Contribution}
\vspace{0.5cm}	

 B.C., D.J.M.\ and S.G.\ contributed to the conception of the experiment, D.J.M., A.T.\ and D.R.A.\ performed the experiment, P.B.\ and S.G.\ made the sample; D.J.M., A.T., D.R.A., B.C., I.A.W.\ and S.G.\ analysed the experimental data and wrote the paper; all authors discussed the results and commented on the paper.

\vspace{0.8cm}
\textbf{Competing financial interests}
\vspace{0.5cm}

The authors declare no competing financial interests.

\end{document}